\documentstyle[12pt]{article}
\def\nsp{\noindent}
\def\eea{\end{eqnarray}}
\def\bea{\begin{eqnarray}}
\def\eeas{\end{eqnarray*}}
\def\beas{\begin{eqnarray*}}
\def\ee{\end{equation}}
\def\be{\begin{equation}}
\def\bdm{\begin{displaymath}}
\def\edm{\end{displaymath}}
\def\fr{\frac}
\def\pa{\partial}

\def\iso{\vec\tau}

\def\vphi{\varphi}
\def\veps{\varepsilon}
\def\thet{\Theta}

\def\thep{\Theta^\prime}
\def\Phip{\Phi^\prime}

\def\dag{^\dagger}
\def\tr{\mbox{Tr}}

\def\fpi2{\mbox{F$_\pi$}^2}

\def\mpi2{{m_\pi}^2}
\def\mk{m_K}
\def\mk2{{m_K}^2}
\def\fkk{\mbox{F$_K$}}
\def\fk2{\mbox{F$_K$}^2}
\def\df{F^\prime}

\def\piv{\hat\pi}
\def\r{\vec r}

\def\omv{\vec\omega}

\def\Fpis{{F_\pi^2}}

\def\Fpis{{F_\pi^2}}
\def\Tr{{\,\mbox{Tr}}}
\def\tr{{\,\mbox{Tr}}}

\def\cai{{\cal I}_i}
\def\caa{{\cal I}_1}
\def\cab{{\cal I}_2}
\def\cac{{\cal I}_3}
\def\cad{{\cal I}_4}

\def\half{\mbox{$\frac{1}{2}$}}
\def\threehalf{\mbox{$\frac{3}{2}$}}
\def\fivehalf{\mbox{$\frac{5}{2}$}}

\begin{document}
\begin{titlepage}
\hfill {\it\today}

\vskip 1.0cm

\begin{center}
{\Large \bf DIBARYONS AS AXIALLY SYMMETRIC\\[1.mm] SKYRMIONS}\\
\vskip 1.4cm
{\large  GILBERTO L. THOMAS}\\
\vskip .3cm
{\large  Instituto de F\'{\i}sica}\\
{\large Universidade Federal do Rio Grande do Sul}\\
{\large  91501--970 Porto Alegre, Brasil}\\
\vskip 0.4cm
{\large NORBERTO N. SCOCCOLA}
\vskip 0.3cm
{\large Physics Dept., Comisi\'on Nac. de Energ\'{\i}a,}\\
{\large Av.Libertador 8250, (1429) Buenos Aires, Argentina.}\\
\vskip 0.4cm
{\large and}\\
\vskip 0.4cm
{\large ANDREAS WIRZBA}\\
\vskip 0.3cm
{\large Institut f\"{u}r Kernphysik}\\
{\large Technische Hochschule Darmstadt}\\
{\large Schlo{\ss}gartenstr.~9, D-W-6100 Darmstadt, Germany}\\
\end{center}

\vskip .4cm

\begin{quotation}

\centerline{\large \bf ABSTRACT}

\vskip 0.3cm

\nsp
{Dibaryons configurations are studied in the framework of the bound state
soliton model. A generalized axially symmetric ansatz is used
to determine the soliton background. We show that once the constraints
imposed by the symmetries of the lowest energy torus configuration
are satisfied all spurious states are removed from the dibaryon
spectrum. In particular, we show that the lowest allowed state in the
$S=-2$ channel carries the quantum numbers of the H particle. We find
that, within our approximations, this particle is slightly bound in the
model. We discuss, however, that vacuum effects neglected in the present
calculation are very likely to unbind the H.}
\end{quotation}

\end{titlepage}

\section{Introduction}

Since it was first proposed by Jaffe \cite{JAF77} that some hexa-quark
state could be stable against strong decays, great attention
has been devoted, both theoretically and experimentally, to this
issue.  Jaffe's suggestion was based on a bag model calculation where
he showed that color-magnetic interactions favor the existence of a
stable flavour singlet $S=-2$ state (the so--called H-dibaryon). Later
on, however, it was shown that the inclusion of effects neglected in
Ref.\cite{JAF77} (like i.e. symmetry breaking effects, center of mass
corrections, pion cloud around the bag, etc.) tend to decrease the
binding in a significant way \cite{OTHERS}, rendering it rather
uncertain.  Moreover, as recently discussed in Ref.\cite{MAL92}, bag
model predictions seem to be very sensitive to the bag constant which
is not strongly constrained by empirical data.  From the experimental
point of view the situation is also unclear.  Although some H weak
decay events have been reported some time ago \cite{SSKM90}, other
recent analysis provides no indication of a stable H--dibaryon
\cite{IMA91}.  New experiments are at the moment being carried out to
investigate this issue further (see i.e. Ref.\cite{QUI92} ).

Theoretical calculations have also been performed using various other
models, like i.e.  lattice QCD, non--relativistic quark model and
soliton models (for a rather extensive list of references, see
Refs.\cite{IMA91,QUI92}).  Again, results have not been conclusive. In
the case of the soliton models, most of the studies have been done
using the collective coordinate $SU(3)$
(see Refs.\cite{BLRS85,KSS92} and references therein)
extensions of
the Skyrme model.  In this paper we will use an alternative method
based on the bound state approach\cite{CK85}. Since it has been
recently shown \cite{RS91,OMRS91} that within this scheme hyperon
properties are remarkably well described, this will provide another
interesting insight into the problem of the strange dibaryons
stability.
A previous attempt to investigate the structure of dibaryons within
the bound state approach was done in Ref.\cite{KM88} where the H was
found to be unbound. In that work, the simplified axially symmetric
ansatz proposed in Ref.\cite{WSH86} was used for the skyrmion field.
Indeed, such a simplified ansatz predicts that the $B=2$ soliton mass
is twice larger than $M_{sol} (B=1)$ and is unstable. Here,
we will use the improved axially symmetric ansatz proposed in
Ref.\cite{KKOS89}.  Although this ansatz does not correspond exactly
to the lowest axially symmetric energy configuration, diskyrmion
properties (like i.e. soliton mass, rotational energies, etc.)
computed with it turn out to be very similar to those obtained with
the lowest energy torus configuration numerically found in
Ref.\cite{KS87}.

An intriguing point discussed in Ref.\cite{KM88} is the existence in
the bound state soliton model of states which are forbidden in the
quark model. In fact, a similar situation was found in
Refs.\cite{WSH86,KKOS89} for the case of non--strange dibaryons. In
this paper we will show, however, that all these spurious states are
removed from the spectrum once the constrains imposed by the
symmetries of the problem are correctly taken into account.  In
particular, in our scheme the lowest $S=-2$ allowed dibaryon state is
the flavour singlet, in agreement with the quark model prediction.

This paper is organized as follows. In Sec.2, we introduce the bound
state model for arbitrary baryon number based on the improved axially
symmetric ansatz. In Sec.3, we show how to construct dibaryon
wavefunctions which are consistent with all the symmetries of the
system.  In Sec.4, numerical results are presented and discussed.
In Sec.5, conclusions are given. In Appendix A, we review the $SU(2)$
sector of the model. Finally, in Appendix B the quantization rules and
the dibaryon quantum numbers are discussed in detail.

\section{The Model}

We start with the effective action for the simple Skyrme model with an
appropriate symmetry breaking term, expressed in terms of the
$SU(3)$--valued chiral field $U(x)$ as
\bea
\Gamma = \int d^4x \ \left\{
{\Fpis \over{16}} \Tr\left[ \partial_\mu U \partial^\mu U^\dagger \right] +
{1\over{32 e^2}} \Tr\left[ [U^\dagger \partial_\mu U ,
U^\dagger \partial_\nu U ]^2 \right] \right\} + \Gamma_{WZ} +
\Gamma_{SB},  \label{action}
\eea
\nsp
where $F_\pi$ is the pion decay constant (~$= 186 \ MeV$ empirically)
and $e$ is the so--called Skyrme parameter.  In Eq.(\ref{action})
$\Gamma_{SB}$ is responsible for the explicit breaking of chiral
symmetry. We use the following form for $\Gamma_{SB}$:
\bea
\Gamma_{SB}
& \! \! = \! \! &\int d^4x \left\{ {\Fpis  m_\pi^2 + 2 F_K^2 m_K^2 \over{48}}
\Tr \left[ U + U^\dagger - 2 \right]
+ { \Fpis m_\pi^2 - F_K^2 m_K^2 \over{24}}
\Tr \left[ \sqrt{3} \lambda^8 \left( U + U^\dagger \right) \right] \right.
\nonumber \\
& & \hskip 1.cm \left. + { F_K^2 - \Fpis \over{48} } \Tr
\left[ \left( 1 - \sqrt{3} \lambda^8 \right)
\left( U \partial_\mu U^\dagger \partial^\mu U +
        U^\dagger \partial_\mu U \partial^\mu U^\dagger \right) \right]
\right\} \ ,
\label{oldmass}
\eea
\noindent
where $\lambda^8$ is the eighth Gell-Mann matrix and $m_\pi$ and $m_K$
represent the pion and kaon masses respectively and $F_K$ is the kaon decay
constant (~$= 1.22 \ F_\pi$). Eq.(\ref{oldmass})
accounts not only for the finite mass of the pseudoscalar mesons but
also for the empirical difference between their decay constants.  In
previous calculations \cite{CK85,KM88}
the kaon was found to be overbound
to the soliton.  It was recently shown \cite{RS91} that this defect
can be mostly eliminated if the difference in the decay constants is
properly taken into account, as done in Eq.(\ref{oldmass}).  Finally,
$\Gamma_{WZ}$ is the Wess--Zumino action,
\be
\Gamma_{WZ} \ = \ -i \fr{N_c}{240\pi^2}\int \ d^5x \
\veps^{\mu\nu\alpha\beta\gamma}
\ \tr(L_\mu L_\nu L_\alpha L_\beta L_\gamma) \ ,
\ee
which distinguishes between states with positive and negative strangeness.

We proceed by introducing the Callan--Klebanov (CK) ansatz for the
chiral field \cite{CK85}
\be
\label{ansatz}
U=\sqrt{U_\pi}U_K\sqrt{U_\pi} \ .
\ee
In this ansatz, $U_K$ is the field that carries
the strangeness. Its form is
\bea
U_K \ = \ \exp \left[ i\fr{2\sqrt2}{{\fkk}} \left( \begin{array}{cc}
                                                        0 & K \\
                                                        K\dag & 0
                                                   \end{array}
                                           \right) \right] \ ,
\eea
where $K$ is the usual kaon isodoublet
\bea
K \ = \ \left( \begin{array}{c}
                   K^+ \\
                   K^0
                \end{array}
                           \right) \ .
\eea

     The other component, $U_\pi$, is the soliton background field. It
is a direct extension to $SU(3)$ of the $SU(2)$ field $u_\pi$, i.e.,
\bea
U_\pi \ = \ \left ( \begin{array}{cc}
                       u_\pi & 0 \\
                       0 & 1
                    \end{array}
                               \right ) \ .
\eea

Replacing the ansatz Eq.(\ref{ansatz}) in the effective action
Eq.(\ref{action}) and expanding up to second order in the kaon
fields we obtain the following lagrangian density for the
kaon--soliton system
\bea
\label{ltot}
  {\cal L} &=& {\cal L}_{SU(2)} +
      (D_\mu K) \dag(D^\mu K) - K^\dagger a_\mu a^\mu  K
\nonumber \\
    &-& \fr{2}{e^2\fk2}\left\{ K\dag K \ \tr\left( [a_\mu, a^\nu]^2 \right)
     - (D_\mu K)\dag(D^\mu K)\mbox{Tr}(a_\nu a^\nu) \right.
\nonumber \\
    &+& \left. (D_\mu K)\dag(D_\nu K)
            \mbox{Tr}(a^\mu a^\nu) - 3(D_\mu K)\dag
                [a^\mu,a^\nu] (D_\nu K) \right \}
\nonumber \\
    &-& i\fr{N_c}{\fk2}B^\mu\left [ K\dag D_\mu K -(D_\mu K)\dag K\right ]
\nonumber \\
    &-& K\dag K \left[ \mk2 - \fr{1}{2}\fr{\fpi2}{\fk2}\mpi2
(1 - \cos F) \right]
\label{kaction}
\eea
\nsp
where ${\cal L}_{SU(2)}$ is the effective pion lagrangian whose explicit
expression is given in the Appendix A and
\bea
D_\mu \ &=& \ \pa_\mu + v_\mu, \nonumber \\
\left( \begin{array}{c}v_\mu \\a_\mu\end{array}\right) \ &=& \
\fr{1}{2}(\sqrt{u_\pi}\dag \pa_\mu \sqrt{u_\pi} \pm
\sqrt{u_\pi} \pa_\mu \sqrt{u_\pi}\dag) \ .
\eea
$B^\mu$ is the baryon number current of the $SU(2)$ configuration
given by
\be
B^\mu \ = \ \fr{1}{24\pi^2} \ \veps^{\mu\alpha\beta\gamma} \
\tr(l_\alpha l_\beta l_\gamma) \ ,
\ee
where $l_\nu = u_\pi\dag \pa_\nu u_\pi$.

In order to obtain the soliton background configuration we introduce
the axially symmetric ansatz
\be
u_\pi = \exp\left[ i \vec \tau \cdot \hat \pi_n \ F \right]
\label{mansatz}
\ee
with
\be
\hat \pi_n = \sin \Theta \cos n\phi \ \hat \imath +
               \sin \Theta \sin n\phi \ \hat \jmath +
               \cos \Theta \ \hat k
\label{pians}
\ee
In Ref.\cite{KM88} it was assumed that $F = F(r)$ and $\Theta =
\theta$, where $(r, \theta, \phi)$ are the usual spherical coordinates.
As already mentioned, such an ansatz, which predicts
\be
R~\equiv~M_{sol}~(B~=~2)~/~M_{sol}~(~B~=~1~)~=~2.14 \ ,
\ee
leads to an unstable soliton configuration. The lowest energy diskyrmion
configuration was numerically found in Ref.\cite{KS87}.
A very good variational approximation to such a solution was proposed in
Ref.\cite{KKOS89}. In this case, $F$ is a function
of $r$ only, but the variational function
\be
\label{eqtheta}
\Theta = \theta + \sum_{k=1}^m g_k \ \sin(2k\theta)
\ee
is used for $\Theta$. The coefficients $g_k$ are determined by
minimizing the soliton energy in the corresponding baryon sector.
Using this ansatz one finds $R=1.94$, which compares
very well with the lowest energy solution value $R_{min} = 1.92$.
As already mentioned, other
computed quantities like rotational energies, baryon radius, etc.\ are
numerically also very close to those of Ref.\cite{KS87}.
Since the use of the variational ansatz leads to a considerable
simplification of the calculation, we will use it to describe the
soliton background.
The expressions
corresponding to the $SU(2)$ sector of the model have been obtained in
Ref.\cite{KKOS89}. For completeness, they are summarized in Appendix
A.

     Of course, for $n=1$ the minimum soliton energy is obtained
for $g_k =0$. Therefore, in this case the background soliton is still
symmetric under combined spatial and isospin rotations
$\vec\Lambda = \vec l + \vec I$ and the kaon field can be expanded
in terms of spinor spherical harmonics,
\be
K(\r,t) \ = \ k(r,t) \ {\cal Y}_{\Lambda l \Lambda_3}(\hat r) \ .
\label{kll}
\ee
However, for $n \ne 1$ the background field is no longer invariant under
$\Lambda$--rotations. In this case we use the
consistent ansatz \cite{KM88}
\be
\label{consatz}
K(\r,t) \ = \ k(r,t) \ \iso\cdot\piv_n(\r) \ \chi \ ,
\ee
where $\chi$ is a two--component spinor. Note that, for $n=1$, this ansatz
reduces to Eq.(\ref{kll}) for the particular
case $\Lambda = 1/2$, $l=1$. These are
precisely the quantum numbers of lowest kaon bound state when $n=1$.

     Using the ans\"atze given above the explicit form of the
kaon-soliton effective lagrangian is
\be
\label{efflag}
L \ = \ \int \ dr \ r^2 \left \{ f(r) \ \dot{k\dag} \dot k - h(r) \
k^{\dagger\prime} k^\prime + i\lambda(r) (\dot{k\dag}k - k\dag \dot k) -
k\dag k(\mk2 + V_{eff}) \right \} \ ,
\ee
where
\be
f(r) \ = \ 1 + \fr{1}{e^2\fk2} \left({\df}^2 + \alpha_1\fr{\sin^2F}{r^2}
\right ) \ ,
\ee
and
\be
h(r) \ = \ 1 + \fr{\alpha_1}{e^2\fk2} \fr{\sin^2F}{r^2} \ .
\ee
The term linear in time derivatives, whose coefficient is
\be
\lambda(r) \ = \ -\fr{\alpha_3 \ N_c}{2\pi^2\fk2} \
\df \ \fr{\sin^2F}{r^2} \ ,
\ee
is due to the Wess--Zumino action, and
\bea
 V_{eff} \ &=& \ \left [ \fr{\alpha_1}{e^2\fk2 r^2} \left ( \cos^4F/2 -
2 \ \sin^2F \right ) - \fr{1}{4}\right ] \ {\df}^2 \nonumber \\
   &-&     \fr{1}{r^2} \left ( \sin^2F - 4 \ \cos^4F/2 \right )
\left ( \fr{2\alpha_2}{e^2\fk2}\fr{\sin^2F}{r^2} +
\fr{\alpha_1}{4} \right )  \nonumber \\
   &-&      3 \fr{\alpha_1}{e^2\fk2 r^2} \fr{d}{dr}
\left [ \df \sin F \ \cos^2 F/2) \right ] -
\fr{1}{2}\fr{\fpi2}{\fk2}\mpi2(1 - \cos F) \ .
\eea
Here, $\alpha_i$ are
\bea
\alpha_1&=&\fr{1}{2} \int_0^\pi d\theta \ \sin\theta
\left( {\thep}^2 + n^2 \ {\sin^2 \thet \over{ \sin^2 \theta} } \right ) \ ,
\nonumber \\
\alpha_2&=&\fr{n^2}{2} \int_0^\pi d\theta
\left( {\thep}^2 \ \fr{\sin^2\thet}{\sin\theta}\right) \ ,\nonumber \\
\alpha_3&=&\fr{n}{2} \int_0^\pi d\theta \ \sin\thet \ \thep \ .
\label{alpha}
\eea

The diagonalization of the hamiltonian obtained from the effective
lagrangian Eq.(\ref{efflag}) leads to the kaon eigenvalue equation
\be
\left[ - {1\over{r^2}} \partial_r \left( r^2 h \partial_r \right)
+ m_K^2 + V_{eff} - f \veps^2  - 2 \ \lambda \ \veps \right] k(r) = 0 \ .
\ee

     To obtain the hyperfine corrections to the dibaryons masses we
proceed with the semiclassical collective coordinates quantization
method, where the isospin and spatial rotations are treated as the
zero modes \cite{ANW83}. Then, we introduce the time--dependent
spatial rotations $R$ and the isospin rotations $A$ such that
\bea
\label{rotpi}
u_\pi \ &\rightarrow & \ R \ A \ u_\pi \ A^{-1} \\
\label{rotk}
K \ &\rightarrow & \ R \ A \ K \ .
\eea
The angular velocities in respect to the body fixed frame are given by
\bea
\label{angvel}
\left ( R^{-1} \dot R \right )_{ab} &=& \veps_{abc}\Omega_c \\
A^{-1} \dot A &=& \fr{i}{2}\ \iso\cdot\omv \ .
\eea
      Using $a_1$ and $a_2$ as coefficients of the up and down spinor
$\chi$ in Eq.(\ref{consatz}) the substitution of
Eqs.(\ref{rotpi},\ref{rotk}) in the full Lagrangian Eq.(\ref{ltot})
yields
\bea
\label{lrot}
L  &=&  - M_{sol} \ + \ L_K \ - \ \vec T\cdot\omv \ +
(T_1\omega_1 +T_2\omega_2 ) s_2 + T_3\omega_3 s_1
\ - \ (T_1\Omega_1 + T_2\Omega_2) t_2 - T_3\Omega_3 t_1  \nonumber \\
&+&  {1\over2} \caa (\Omega^2_1+\Omega^2_2) \ + \ {1\over2} \cab
(\omega^2_1+\omega^2_2) \ - \
\cad \delta_{n,1}(\Omega_1\omega_1 + \Omega_2\omega_2) \ + \
\fr{1}{2}\cac (n\Omega_3-\omega_3)^2 \ ,
\eea
where \footnote{In what follows we assume that $\Theta$ = $\theta$ for
$n = 1$.}
\bea
s_1  \ & = & \ 3 \left \{ 2 \alpha_4 \ d_1 \ + \
\left [ \alpha_5 - \alpha_1 \alpha_4) \right ] \ d_2 \right \} \ , \\
s_2  \ & = & \ 3 \left \{ ( 1- \alpha_4 ) \ d_1 \ + \ \fr{1}{2}
\left [ \alpha_1 (1+\alpha_4) - \alpha_5) \right ] \ d_2 \right \} \ , \\
t_1  \ & = & \ n \ s_2 \ , \\
t_2 \ & = & \ 2 ( d_1 \ + \ d_2 ) \ \delta_{n,1} \ ,
\eea
with $d_1$ and $d_2$ given by
\bea
d_1 \ &=& \ 2\veps_n \ \int_0^\infty \ dr \ k^*k \ \left[ \fr{2}{3}r^2 f
\cos^2F/2 \ -
\ \fr{1}{e^2\fk2}\fr{d}{dr} (r^2 \ \df \sin F) \right ] \\
d_2 \ &=& \ \fr{2\veps_n}{e^2\fk2} \ \int_0^\infty \ dr \ k^*k \;
\fr{4}{3}\cos^2F/2 \
\sin^2 F \ ,
\eea
and the angular integrals $\alpha_4$ and $\alpha_5$ by
\bea
\alpha_4 \ &=& \ \fr{1}{4}\int_0^\pi \ d\theta \ \sin\theta \ \sin^2\thet \ ,
\\
\alpha_5 \ &=& \ \int_0^\pi \ d\theta \ \sin\theta \ \sin^2\thet \
{\thep}^2 \ .
\eea

In Eq.(\ref{lrot}) $T^l$ is defined as $T^l=a^*_i\tau^l_{ij}a_j$ and
$\cai$ are the moments of inertia of $SU(2)$ sector whose explicit
expressions are given in Appendix A.  The spin and isospin components
$J^{bf}_i$ and $I^{bf}_i$ respectively, are calculated via
\be
J^{bf}_i \ = \ \fr{\pa L}{\pa\Omega_i} \ , \ \ \ \ \ \ \ \ \ \ \ \ I^{bf}_i \
= \ \fr{\pa L}{\pa\omega_i} \ .
\ee
Unlike the $n=1$ case where the hedgehog symmetry of the skyrmion
enforces the constraint $\vec J + \vec I = \vec T$, for $ n \ne 1 $
only the constraint
\be
J^{bf}_3 \ = \ - n (I^{bf}_3 \ + \ T_3)
\label{constraint}
\ee
due to axial symmetry is left.

The quantization of the rotational hamiltonian leads to
\bea
H_{rot} \ &=& \ \fr{1}{2\caa}(\vec J^2-(J_3^{bf})^2) \
+ \ \fr{1}{2\cab}(\vec I^2- (I_3^{bf})^2) \ +
\ \fr{c^2_2}{2\cab}(\vec T^2-T_3^2) \nonumber \\
&+& \ \fr{1}{2\cac}(I_3^{bf}+c_1 T_3)^2 \ +
\ \fr{c_2}{2\cab}(I_+^{bf} T_- + I^{bf}_- T_+) \
,
\label{hrot}
\eea
with the hyperfine constants $c_1$ and $c_2$ given by
\bea
c_1 \ &=& \ 1 \ - \ s_1  \\
c_2 \ &=& \ 1 \ - \ s_2  \ .
\eea

Given this form of the rotational hamiltonian we should find the
corresponding eigenfunctions that satisfy the constraints imposed
by the symmetries of the system. This is done in the following
section.

\section{Dibaryon wavefunctions}

In general the dibaryon wave functions
are combinations of product states of the rotation matrices
$D^I_{I_3,I_3^{bf}} (\omega)$ for isospin, $D^J_{J_3,J_3^{bf}}
(\Omega)$ for the angular momentum and the kaon eigenstates $k_{T_3}
(\vec r, t)$. Here $I_3$ and $J_3$ are respectively the isospin and
angular momentum projection on the lab.  frame, $I_3^{bf}$ and
$J_3^{bf}$ on the body fixed frame and $T_3$ the projection of the
kaon "spin" on the body fixed frame. Each of these product states
have the form
\be
D^J_{J_3, J_3^{bf}} (\Omega) \
D^I_{I_3, I_3^{bf}} (\omega) \
k_{T_3} (\vec r, t)
\label{wf1}
\ee
where
\be
J_3^{bf} = -2 (I_3^{bf} + T_3^{bf}).
\label{axialconstraint}
\ee

First we note that in these product states, for a given value of $S$, not
all the values of isospin are allowed. This can be shown using the same
method used in Ref.\cite{SW91} for $B=1$. Details are given in Appendix B.
One obtains that for states in the minimal
representations the allowed values of isospin are given by
\be
I = {p\over2}
\label{isoconstraint}
\ee
where $p\le 6+S$ should be odd if $S$ is odd or even otherwise.
This relation together with the axial symmetry constraint
Eq.(\ref{axialconstraint}) imply that $J_3^{bf}$ is always even.

To determine for each set of quantum numbers
the correct linear combination of product
states we should take into
account all the symmetries of the problem. As discussed in
Ref.\cite{KKOS89} in addition to axial symmetry, the ansatz
Eqs.(\ref{pians},\ref{eqtheta}) has, for $n=2$, the following
reflection symmetries
\bea
\hat \pi_2 (-x,y,z) &=& \left( \begin{array}{rrr} 1 &    &  \\
                                                    & -1 &  \\
                                                    &    & 1
                                \end{array} \right)
                         \hat \pi_2 (x,y,z) \label{sim1}       \\
\hat \pi_2 (x,-y,z) &=& \left( \begin{array}{rrr} 1 &    &  \\
                                                    & -1 &  \\
                                                    &    & 1
                                \end{array} \right)
                         \hat \pi_2 (x,y,z) \label{sim2}       \\
\hat \pi_2 (x,y,-z) &=& \left( \begin{array}{rrr} 1 &    &  \\
                                                    &  1 &  \\
                                                    &    & -1
                                \end{array} \right)
                         \hat \pi_2 (x,y,z) \label{sim3}
\eea

Denoting spatial rotations through an angle $\phi$ about the
$i$-axis by ${\cal R}_i^J (\phi)$ and isorotations through
an angle $\theta$ about the $j$-axis by ${\cal R}_j^I (\theta)$
the symmetry transformations are also generated by
\bea
& &{\cal{R}}_1^J(\pi) \ {\cal{R}}_1^I(\pi)
\label{simopone} \\
& &{\cal{R}}_2^J(\pi) \ {\cal{R}}_1^I(\pi)
\label{simoptwo} \\
& &{\cal{P}} \ {\cal{R}}_3^I(\pi)
\label{simopthree}
\eea
The operator ${\cal P}$ denotes the parity operator which is defined
as a space inversion and a sign change of all components of the pion field.

The importance of the first two of these symmetry transformations on
the construction of the dibaryon wavefunctions was first
noticed in Ref.\cite{BC86}. In fact, as it discussed in
detail in Sec.4-2c of Ref.\cite{BM75}, for system with axial
symmetry, the existence of an additional symmetry
with respect to a $\pi$-rotation about an axis
perpendicular to the symmetry axis implies that this $\pi$-rotation
is part of the intrinsic degrees of freedom, and is therefore
not to be included in the rotational degrees of freedom.
We can express this constraint by requiring that
the operator ${\cal R}_{ext}$, which performs a rotation ${\cal R}$
by acting on the collective orientation angles (external variables),
is identical to the operator ${\cal R}_{int}$, which performs the
same rotation by acting on the intrinsic variables,
\be
{\cal R}_{ext} = {\cal R}_{int} \ .
\label{pirotconstraint}
\ee
Since in our case
${\cal R}$ can be either given by Eq.(\ref{simopone}) or
Eq.(\ref{simoptwo}) we will have, in principle, two independent
constraints.
We will see, however, that when applied
to the product states both symmetry operations produce
the same result and therefore we are left with one single
constraint.

Then, in order to satisfy the constraints imposed by
the symmetries of the ansatz we have to choose the following
linear combination of product states:
\be
N \left[ 1 + ({\cal R}_{ext})^{-1}  {\cal R}_{int} \right]
D^J_{J_3, J_3^{bf}} (\Omega) \
D^I_{I_3, I_3^{bf}} (\omega) \
k_{T_3} (\vec r, t) \ ,
\label{lincomb}
\ee
where $N$ is a normalization factor. In writing Eq.(\ref{lincomb})
we have use the fact that acting on product states
$({\cal{R}}_{ext})^2 = ({\cal{R}}_{int})^2$.
As we will see below this is satisfied by our product states.

In order to have the explicit form of the dibaryon wave
function we should apply the symmetry operators on the product
wave function. For the collective operators we get
\bea
\left[ {\cal{R}}_1^J(\pi) {\cal{R}}_1^I(\pi) \right]^{-1}
D^J_{J_3, J_3^{bf}} (\Omega) D^I_{I_3, I_3^{bf}} (\omega)
&=& (-)^{I+J}
D^J_{J_3, - J_3^{bf}} (\Omega) D^I_{I_3, - I_3^{bf}} (\omega) \ ,
\label{colopone}\\
\left[ {\cal{R}}_2^J(\pi) {\cal{R}}_1^I(\pi) \right]^{-1}
D^J_{J_3, J_3^{bf}} (\Omega) D^I_{I_3, I_3^{bf}} (\omega)
&=& (-)^{I+J-J_3^{bf}}
D^J_{J_3, - J_3^{bf}} (\Omega) D^I_{I_3, - I_3^{bf}} (\omega) \ .
\label{coloptwo}
\eea

To determine the effect of the symmetry transformations
on the intrinsic wave function we have to notice that
apart from the contribution of the kaon field we have to
include the contribution of the soliton itself. The latter
contribution has been calculated in Refs.\cite{BC86,VWWW87}.
Using the fermionic nature of the $B=1$ soliton configuration
they found\footnote{Note that when one extends the light flavour group
from $SU(2)$ to $SU(3)^f$ as done in Ref.\cite{SW91} the minus phase is
obtained from the Wess-Zumino term.}
\bea
{\cal{R}}_1^J(\pi) {\cal{R}}_1^I(\pi) \ \psi_{sol} &=& - \ \psi_{sol} \ ,
\label{solopone} \\
{\cal{R}}_2^J(\pi) {\cal{R}}_1^I(\pi) \ \psi_{sol} &=& - \ \psi_{sol} \ .
\label{soloptwo}
\eea
Finally, we have to calculate the effect of the symmetry transformations
on the kaon field. Performing the symmetry operations on
one kaon states we find
\bea
{\cal{R}}_1^J(\pi) {\cal{R}}_1^I(\pi) \ k_{T_3} (S= \pm 1) &=&
\mp i \ k_{-T_3} (S= \pm 1) \ ,
\label{okopone} \\
{\cal{R}}_2^J(\pi) {\cal{R}}_1^I(\pi) \ k_{T_3} (S= \pm 1) &=&
\mp i \ k_{ -T_3} (S= \pm 1) \ .
\label{okoptwo}
\eea
In deriving Eqs.(\ref{okopone},\ref{okoptwo}), we have use
the form Eq.(\ref{consatz})
for $S=+1$ kaons and its charge
conjugate for $S=-1$ kaons. Therefore, for a state with arbitrary
value of strangeness $S$ we obtain
\bea
{\cal{R}}_1^J(\pi) {\cal{R}}_1^I(\pi) \ k_{T_3} &=&
(-)^{-S/2} k_{-T_3} \ ,
\label{kaonopone}  \\
{\cal{R}}_1^J(\pi) {\cal{R}}_1^I(\pi) \ k_{T_3} &=&
(-)^{-S/2} k_{ -T_3} \ .
\label{kaonoptwo}
\eea

As we see, the only difference between the action of the symmetry
transformation Eq.(\ref{simopone}) and that of Eq.(\ref{simoptwo})
is the presence of an extra phase $(-)^{J_3^{bf}}$ in Eq.(\ref{coloptwo}).
Since we have
shown that in our case $J_3^{bf}$ is always even, we see
that, as stated above, both symmetry
operations produce that same effect when
acting on our product states. Moreover, using
the constraints Eqs.(\ref{axialconstraint},\ref{isoconstraint})
together with the effect of the symmetry operations on the
intrinsic and collective wave functions
Eqs.(\ref{colopone}-\ref{kaonoptwo}),
it is easy to show that ${\cal R}_{ext}^2 = {\cal R}_{int}^2$ when
applied to the product states.

Therefore, the dibaryon wave functions that satisfy all the
constraints imposed by the symmetries of the system have
the structure
\bea
| I I_3, J J_3, S > &=& N \left(
D^J_{J_3, -2K } (\Omega)
D^I_{I_3, K - T_3} (\omega)
k_{T_3} (\vec r, t) - \right.
\nonumber \\
& & \qquad \qquad \left. (-)^{I+J-S/2}
D^J_{J_3,  2 K } (\Omega)
D^I_{I_3, - K + T_3} (\omega)
k_{-T_3} (\vec r, t)
\right) \ ,
\label{bthirteen}
\eea
where $K = I_3^{bf} + T_3$. The normalization constant $N$ can be
easily calculated. We obtain
\be
N \ = \ \fr{1}{\sqrt{2(1+\delta_{I_3^{bf},0} \delta_{T_3,0})}}
\fr{\sqrt{(2J+1)(2I+1)}}{8\pi^2} \ .
\ee

To determine the parity of these wavefunctions we have to make use of
the symmetry operation given in Eq.(\ref{simopthree}). In fact,
as discussed in Sec.4-2f of Ref.\cite{BM75} the existence of such
kind of symmetries implies that the parity operator ${\cal P}$ can
be written as
\be
{\cal P} = {\cal S} \ {\cal R}^{-1} \ ,
\ee
where ${\cal S}$ (= ${\cal P} {\cal R}_3^I (\pi)$ in our case) acts
on the intrinsic coordinates, while ${\cal R}^{-1}$ ( =
$[{\cal R}_3^I (\pi)]^{-1}$ in our case ) acts on the collective
variables. Using the fact that $K$ is always integer and that
\bea
{\cal S} \ \psi_{sol} &=& \psi_{sol}  \\
{\cal S} \ k_{\pm T_3} &=& (-)^{\pm T_3} \ k_{\pm T_3} \\
{\cal R}_3^I (\pi) \ D^J_{J_3, \mp 2K } (\Omega)
D^I_{I_3, \pm (K - T_3)}
(\omega) &=& (-)^{\pm (K-T_3) } \ D^J_{J_3, \mp 2K } (\Omega)
D^I_{I_3, \pm (K - T_3)} (\omega)
\eea
we find that the parity of the wavefunction Eq.(\ref{bthirteen})
is given by
\be
\pi = (-)^K.
\label{parity}
\ee

 From the explicit form of the wavefunction we note
that for $K=0$, $T_3=0$ only the ``Fermi--allowed" combinations
(satisfying the constraint $(-)^{I+J-S/2} = -1$ ) survive, whereas the
``Fermi--forbidden" ones have zero norm. For the particular
case of the two-nucleon system ($S=0$), we have the constraint
$(-)^{I+J}=-1$ which is nothing but the generalized form of the
Pauli principle. It is interesting to note that in our case
this principle realizes in a more conventional way than
in Ref.\cite{VWWW87} (see Eqs.(5.7a-b) of that reference).
This is due to the more convenient choice of spatial coordinates
we have used in the present work.
In general, using the constraint Eq.(\ref{isoconstraint})
together with the explicit form of the dibaryon eigenstates
Eq.(\ref{bthirteen}), one readily finds that all the spurious states
obtained in Refs.\cite{KM88,WSH86,KKOS89} are removed from the spectrum.
The quantum numbers of the allowed states with $J \le 2$ are shown in
Table 1.

An important remark is that the wavefunction obtained above is not an
eigenfunction of the rotational hamiltonian $H_{rot}$.
Since the Coriolis term (last term in Eq.(\ref{hrot}))
does not commute with $T_3$, the
eigenfunctions of the $H_{rot}$ will be combinations of those
given in Eq.(\ref{bthirteen}) with the same $I I_3, J J_3,
(J_3^{bf})^2$ quantum numbers, but different values of $T_3^2$.

Using Eqs.(\ref{hrot},\ref{bthirteen}) one can finally obtain the
dibaryon mass formula. For ground state dibaryons we get
\be
M = M_{sol} + | S | \ \veps + M_{rot}(S)
\label{massformula}
\ee
where $S$ is the strangeness of the state and $M_{rot}(S)$ is the
rotational contribution. Since for $S=0$ and $S=-1$ only one value
of $T_3^2$ is allowed ($T_3^2 = 0$ for $S=0$ and
$T_3^2 = 1/4$ for $S=-1$),
in these cases $M_{rot}$ is simply given by the mean value of
$H_{rot}$ in the corresponding dibaryon state.
For $S \ = \ 0$ we get
\be
M_{rot}^{S=0} = \fr{1}{2\caa}
\left [ J (J+1) \ - \ (J_3^{bf})^2 \right ] \ + \
\fr{1}{2\cab} \
\left [ I (I+1) \ - \ (I_3^{bf})^2 \right ] \ + \ \fr{1}{2\cac}
(I_3^{bf})^2 \ ,
\ee
i.e., the same as \cite{WSH86}.

     For $S \ = \ -1$, the rotational contribution is given by
\bea
M_{rot}^{S=-1} &=& \fr{1}{2\caa} \left [ J (J+1)  -  (J_3^{bf})^2 \right ]
+  \fr{1}{2\cac} \left \{ (1 - c_1) \left [ (I^{bf}_3)^2 \ - \
\fr{c_1}{4}\right
] \ +  \fr{c_1 (J_3^{bf})^2}{4} \right \} \nonumber \\
&+& \fr{1}{2\cab} \left [ I (I+1)  -  (I_3^{bf})^2  +  c_2
\left ( \fr{c_2}{2} - (-)^{I+J+1/2} \delta_{J_3,0} \sqrt{ I(I+1) + 1/4 }
\right ) \right ] \ .
\eea

For $S \ = \ -2$, two values of $T_3^2$ ( = 0, 1) are in general
allowed \footnote{
Whether {\it both} values are allowed
for a given state depends on the values of the other quantum
numbers.}.
Therefore, $M_{rot}$ is given by the eigenvalues of
\bea
M_{rot}^{S=-2} & = & \ \fr{1}{2\caa} \left [ J (J+1) \ - \ (J_3^{bf})^2 \right
]
\ + \ \fr{1}{2\cab} \ \left [ I (I+1) \ + \ c_2^2 \right ] \ + \
\fr{c_1 K^2}{2\cac} \nonumber \\
&   & \!\!\!\!\!\!\!\!\!\!\!\!\!\!\!\!\!\! + \left ( \begin{array}{cc}
\fr{c^2_2 - K^2}{2\cab} + \fr{1 - c_1}{2\cac} K^2 & \fr{c_2}{\cab\sqrt{2}}
\sqrt{1 + \delta_{K,0}} \sqrt{(I \mp K)(I \pm M)} \\
\fr{c_2}{\cab\sqrt{2}} \sqrt{1 + \delta_{K,0}} \sqrt{(I \mp K)(I \pm M)} &
\fr{1 - c_1}{2\cac}\left ( M^2 - c_1 \right ) - \fr{M^2}{2\cab}
\end{array}
\right )
\label{mrotst}
\eea
with $M=K \pm 1$. In the construction of
2 x 2 matrix we have used the basis given in Eq.(\ref{bthirteen})
ordered according to increasing $T_3^2$.

Note that although in some particular cases the
expressions for the rotational corrections to
$S=-1$ and $S-2$ dibaryons agree with those given
in Ref.\cite{KM88}, in general they differ.
This is due to the fact that in general the eigenfunctions used
in that reference do not satisfy the constraints imposed
by the symmetry transformations Eqs.(\ref{simopone},\ref{simoptwo}).

\section{Results and discussion}

In our numerical calculations we will consider two sets of values for
the parameters in the effective action. In one case (SET A), we
consider the chiral limit in the $SU(2)$ sector, $m_\pi=0$ and fit
$F_\pi$ and $e$ to reproduce the empirical $N$ and $\Delta$ masses.
This corresponds to the result of Ref.\cite{ANW83},
\be
F_\pi = 129 \ MeV, \ \ \ \ \ \ \ e =5.45 \ .
\ee
The second set of parameters (SET B) is obtained for $m_\pi = 138 \
MeV$.  It corresponds to the result of Ref.\cite{AN84}
\be
F_\pi = 108 \ MeV, \ \ \ \ \ \ \ e =4.84 \ .
\ee
In both cases
we take the kaon mass and the ratio $F_K/F_\pi$ to their empirical
values $m_K = 495 \ MeV$, $F_K/F_\pi = 1.22$. When comparing our
results with those of Ref.\cite{KM88}, it should be noticed that in
that reference the ratio of decay constants was taken to be 1. As
mentioned above this leads to an important overbinding in the $B=1$
sector. The spectrum of non-strange and strange baryons obtained for
our two parameters sets can be found for example in Ref.\cite{RS91}
and it will be not repeated here.

As shown in Ref.\cite{KKOS89} when SET B is used and only $g_1$ is taken
as variational parameter, the minimum in the $B=2$ soliton mass
is found for
\be
g_1 = -0.339    \ \ \ \ \ g_{i\ne1} = 0 \ .
\label{gi}
\ee
The inclusion of $g_2$ as a second variational parameter does not
lead to any significant improvement in the energy minimum. In addition
we have also checked that for SET~A the minimum appears almost at
the same value of $g_1$. For this reason we will use the set of
$g_i$ given in (\ref{gi}) in all our calculations.

In the Table 2, we present the calculated values of all the parameters
appearing in the dibaryon mass formulae for both the massless and the
massive pion cases. Non-strange sector parameters corresponding to SET B
have already been given in Ref.\cite{KKOS89}. They are repeated here for
completeness. The values of $\veps$ indicate that kaons are less
bound to the soliton than in the $B=1$ case. Using the
$B=1$ kaon eigenenergies given in Ref.\cite{OMRS91}, we get
\be
\Delta \veps \equiv \veps (B=2) - \veps (B=1) = 16 \ MeV
\ee
for both, SET~A and SET~B. A similar result was found in
Ref.\cite{KM88}, although in that case the value of $\Delta \veps$ was
somewhat smaller, $\Delta \veps \simeq 10 \ MeV$.  It should be
noticed that when the empirical value of the meson decay constant
ratio is used in order to eliminate the large kaon overbinding found
in Ref.\cite{KM88}, $\Delta \veps$ becomes smaller. In fact, if we set
all $g_i=0$ and use SET~A as done in Ref.\cite{KM88} but keep
$F_K/F_\pi = 1.22$, we find $\Delta \veps = 0$. On the other hand, the
use of the improved axially symmetric ansatz increases the value of
$\Delta \veps$, the effect being larger for $F_K/F_\pi = 1.22$ (=16
$MeV$) than for $F_K/F_\pi = 1$ (= 8 $MeV$). Similar comments hold for
SET~B.

The calculated rotational corrections $M_{rot}$ to the dibaryon masses
are shown in Table 3. We list only the allowed states with $M_{rot}
\le 250 \ MeV$. For comparison, we also give the results of
Ref.\cite{KM88} and the sum of the rotational contribution to the
lowest baryon-baryon state in each particular channel. We observe that
our rotational energies are somewhat dependent on the choice of
parameters, being smaller for SET B. For SET A they are in general
very similar to those reported in Ref.\cite{KM88}. An important
exception is the lowest $S=-2$ state for which we predict roughly half
of the value given there. The reason for this difference is mainly the
smaller value of $c_2$ obtained in our model.  This is of some
importance since in our calculation the rotational energy of this
state lies below the corresponding threshold for both sets of
parameters in contrast with the situation in Ref.\cite{KM88}.
Another interesting
$S=-2$ state to be discussed is the $(1,1^+)$ state. This is lowest
state for which the off-diagonal terms in Eq.(\ref{mrotst}) do not
vanish. Therefore, in order to calculate the corresponding rotational
energy the 2~x~2 matrix has been diagonalized.  It should be noticed
than in doing so one goes beyond the ${\cal O} (N_c^{-1})$ in the
$1/N_c$ expansion (where $N_c$ is the number of colors)\footnote{For a
similar situation in a slightly different context see
Ref.\cite{OMRS91}}. Strictly speaking this is not consistent with the
fact that other contributions to the same order have been
systematically neglected.  However, if we just consider the
corresponding diagonal term in Eq.(\ref{mrotst}) as the rotational
correction to such an state one obtains $M_{rot}(1,1^+) = 154 \ MeV$
(SET A). This rather large rotational correction is in disagreement with
different calculations (see i.e. Ref.\cite{KSS92}) that predict the
existence of a low-lying state with these quantum numbers.

Now we will focus our attention on the problem of the H stability.
The different contributions to the corresponding
binding energy are given in Table 4. As we see, for both
sets of parameters, the H is bound in our model.
Since the attraction found
at the level of hyperfine (rotational) interactions is not
enough to compensate the fact that kaons are less bound
to the diskyrmion, the main source of attraction turns out to
be the rather low value of the $B=2$ soliton mass obtained with
the improved axially symmetric ansatz.
A similar situation occurs in soliton models calculation
based on the $SU(3)$ collective coordinates approach.
At this point it is important to recall that although the lowest energy
state in the $S=0$ channel has the quantum numbers of the deuteron its
identification with this particle it is not completely clear. In particular,
a large binding ( $\simeq 130 MeV$ ) is predicted for
this state in the present model \cite{KKOS89}. Since a considerable part
of this overbinding comes from the ${\cal O}(N_c)$ contribution to the
dibaryon mass this might be an indication that the static axial symmetric
torus does not provide the correct description of the $B=2$ baryons.
In using this configuration only the potential
energy aspect of the dibaryon state is addressed, but all of the
relative kinetic energy contribution has been neglected.
In fact, it has been recently shown in Ref.\cite{SRY93}
where a bound two-skyrmion configuration was numerically studied on a
discrete mesh, the torus state was formed only at the closest encounter
of the two skyrmions, whereas most of the time the two-skyrmion system
consisted out of two well separated baryons as expected for a
``well-respecting" deuteron. Another effect that has been ignored
in our calculation (and which is possible related with the lack of
kinetic corrections mentioned above) is the Casimir effect due to
zero point vibrations. A recent
estimation of this effect using the $SO(3)$ ansatz \cite{SSG93} indicates
that it tends to shift the mass of the $B=2$ configuration
relative to two $B=1$
hedgehogs upwards. If we assume that this is the main source of repulsion
that brings the deuteron mass to its physical value, then such a repulsion
would be enough to unbind the H-dibaryon.

\section{Conclusions}

In this article we have studied the structure of strange dibaryons
in the context of the bound state soliton model. In this approach,
such dibaryons are assumed to be bound states of kaons and a $B=2$
topological soliton. To describe the diskyrmion configuration we have used
an axially symmetric ansatz where the dependence on the azimuthal angle
$\theta$ is found through a variational method. Such an ansatz provides
a very good approximation to the numerically found lowest energy
solution which also has axial symmetry. We have shown that once the
constrains imposed by the symmetries of the torus background
configuration are satisfied all spurious states are eliminated
from the spectrum. In particular, we obtained a generalized
form of the Pauli principle for the lowest lying $B=2$ states
with even values of strangeness. In the $S=0$ sector this implies
that the lowest allowed state has the quantum numbers of the deuteron,
while for the $S=-2$ sector the lowest state has the quantum numbers
of the H particle. This is in agreement with the predictions of
the quark based models.

We have found that although kaons are less bound to the diskyrmion
configuration than to a single soliton, the
H-dibaryon is barely bound within our approximations.
This binding is mainly due to the rather small diskyrmion mass
with respect to two individual skyrmions.
However, we know from the case of the deuteron (which is strongly
overbound in the present model\cite{KKOS89})
that the static torus configuration tends to underestimate the $B=2$
soliton mass. Moreover, numerical studies \cite{SRY93} showed that
the torus configuration is formed only at the closest encounter of two
skyrmions. In this sense it would be interesting to see whether the
dynamical departures from the lowest energy solution can be parametrized
in terms of some collective coordinate. Zero-point fluctuations
of this coordinate could then provide
a mechanism to increase the $B=2$ soliton mass without affecting
the $B=1$ skyrmion mass. Another effect that is expected to decrease
the binding energies (and which is probably very much related with the
previous one) is the Casimir effect which leads to contributions
of ${\cal O} (N_c^0)$ \cite{SSG93}. Since these effects are expected to
give similar contributions to all the dibaryon
states independently of their strangeness, it is clear that if they are
strong enough to push the deuteron mass to its empirical value
they are very likely to unbind the H.

\section*{Acknowledgements}

NNS would like to thank D.R. Bes and R.De Luca for very
enlighting discussions.
GLT would like to thank FINEP and FAPERGS (Brazil) for
financial support. Two of us (NNS and AW) would like to thank
the Niels Bohr Institute and Nordita at Copenhagen for their
warm hospitality at some stage of the present work.

\newpage

\section*{Appendix A: The $SU(2)$ Sector}

\renewcommand{\theequation}{A.\arabic{equation}}
\setcounter{equation}{0}

In this appendix, we give, for completeness the expressions for the
$SU(2)$ sector of our model. Most of them have been already published
elsewhere \cite{KKOS89}.

     The lagrangian density ${\cal L}_{SU(2)}$ in Eq.(\ref{ltot})
is
\bea
{\cal L}_{SU(2)}& = &
     \fr{\fpi2}{16} \tr(\pa_\mu{u_\pi}\dag \pa^\mu u_\pi)+
     \fr{1}{32e^2} \mbox{Tr} \left([\pa_\mu u_\pi {u_\pi}\dag,
     \pa_\nu u_\pi{u_\pi}\dag][\pa^\mu u_\pi{u_\pi}\dag,
     \pa^\nu u_\pi{u_\pi}\dag] \right)
\nonumber \\
    & & - \fr{1}{4} \fpi2 \mpi2 (1 - \cos F) \ .
\eea

Replacing the modified ansatz given in Eq.(\ref{mansatz}) the
classical mass turns out to be a functional of $F$, $\thet$ and $\Phi$
which explicit expression is

\bea
\label{msol}
M[F,\thet,\Phi] \ &=& \ \int \ d^3\r \ \left \{ \fr{\fpi2}{8} \left [ {\df}^2
+
\left ( {\thep}^2 + \fr{\sin^2\thet}{\sin^2\theta} \ {\Phip}^2 \right )
\fr{\sin^2F}{r^2} \right ] \right. \nonumber \\
&+& \fr{1}{2e^2}\fr{\sin^2F}{r^2} \left [ \left ( {\thep}^2 +
\fr{\sin^2\thet}{\sin^2\theta} \ {\Phip}^2 \right ) \ {\df}^2 \right. \nonumber
\\
&+& \left. \left. \fr{\sin^2\thet}{\sin^2\theta} \ {\thep}^2 \ {\Phip}^2 \
\fr{\sin^2F}{r^2} \right ] + \fr{\mpi2\fpi2}{4}(1 - \cos F) \right \} \ ,
\eea
with
\bea
\df \ = \ \fr{dF}{dr} \ , \ \ \ \ \thep \ = \ \fr{d\thet}{d\theta} \ \ \ \ \
\mbox{and} \ \ \ \ \Phip \ = \ \fr{d\Phi}{d\phi} \ .
\eea

The minimization of Eq.(\ref{msol}) leads to an Euler--Lagrange for
each of the functions $F$, $\thet$ and $\Phi$.
The first, and more simple, is
\be
\Phi^{\prime\prime} = 0 \ .
\ee
Due to single--valueness of the chiral field we must have $\Phi = n\vphi$,
with $n$ the baryon number. Using this result for $\Phi$, the equation for
$\thet$ is
\bea
\label{theta}
\lefteqn{\left ( C_1\sin\theta + C_2\fr{n^2}{\sin\theta}\sin^2\thet \right
)\thet^{\prime\prime} + \left ( C_1 - C_2n^2\fr{\sin^2\thet}
{\sin^2\theta} \right)\cos\theta \ \thep } \nonumber \\
& - & \left (C_1 - C_2{\thep}^2\right ) n^2
\fr{\sin\thet}{\sin\theta}\cos\thet = 0 \ ,
\eea
with
\be
C_1 \ = \ 2\pi\int_0^\infty dr \ \sin^2F \ \left ( \fr{\fpi2}{8} + \fr{1}{2e^2}
{\df}^2 \right ) \ \ \ \ \mbox{and} \ \ \ \ C_2 \ = \ 2\pi \int_0^\infty dr \
\fr{1}{2e^2}\fr{\sin^4F}{r^2} \ .
\ee
As pointed out by Kurihara et al. \cite{KKOS89},
if one uses the form $\thet=\theta$ as
done in Ref.\cite{WSH86} this equation implies
\be
C_1(1 - n^2)\cos\theta \ = \ 0 \ .
\ee
which can only be satisfied for $n=\pm 1$.
Therefore, for $n\neq \pm 1$ there is a local
unstability. Instead of solving Eq.(\ref{theta}) numerically
Kurihara {\it et al.} proposed to use the trial function defined in
Eq.(\ref{eqtheta}).

     Finally, the equation for the chiral angle $F$ is
\bea
\label{Feq}
\left ( \fr{\fpi2}{4}r^2 + \fr{\alpha_1}{e^2}\sin^2F \right ) \
F^{\prime\prime} &+& \fr{\fpi2}{2} \ r \ \df - \fr{\fpi2}{4}\alpha_1
\sin F \cos F + \fr{\alpha_1}{e^2} {\df}^2 \sin F \cos F  \nonumber \\
& & \!\!\!\!\!\!\!\!\!\!\!\!\!\!\!\!\!\!\!\!\!\!\!\!\!\!\!\!\!\!\!
- \ 2 \fr{\alpha_2}{e^2} \fr{\sin^3F\cos F}{r^2} -
\fr{\mpi2\fpi2}{4}r^2 \sin F = 0 \ ,
\eea
with the boundary conditions $F(0)=\pi$ and $F(\infty)=0$. The explicit
expressions of $\alpha_1$ and $\alpha_2$ have been given in
Eq.(\ref{alpha}).

     The use of the collective coordinate method for the quantization of
the $SU(2)$ Lagrangian leads to the four last terms of Eq.(\ref{lrot}).
The explicit form of the moments of inertia $\cai$ appearing in such an
equation is
\be
{\cal I}_i \ = \ \int_0^\infty \ dr \ r^2 \sin^2F \left [ \left ( \fr{\fpi2}{4}
\ + \ \fr{1}{e^2}{\df}^2 \right ) \zeta_i \ + \ \fr{1}{e^2}
\fr{\sin^2F}{r^2}\eta_i \right ] \ ,
\ee
where the coefficients $\zeta_i$ and $\eta_i$ are given by
\bea
\begin{array}{ll}
\zeta_1 = \pi\int_0^\pi \ d\theta \ \sin\theta \left ( {\thep}^2 + n^2
\fr{\sin^2\thet}{\tan^2\theta} \right ) \ , & \eta_1 = n^2 \pi\int_0^\pi \
d\theta
\ \sin\theta (1 + \cos^2\theta) {\thep}^2 \ \fr{\sin^2\thet}{\sin^2\theta} \ ,
\\
\zeta_2 = \pi
\int_0^\pi \ d\theta \ \sin\theta (1 + \cos^2\thet) & \eta_2 = \pi
\int_0^\pi \ d\theta \ \sin\theta \left ({\thep}^2\cos^2\thet + n^2
\fr{\sin^2\thet}{\sin^2\theta} \right ) \ , \\
\zeta_3 = 2\pi\int_0^\pi \ d\theta \ \sin\theta \ \sin^2\thet & \eta_3
= 2\pi\int_0^\pi \ d\theta \ \sin\theta \ {\thep}^2\sin^2\thet \ , \\
\zeta_4 = \fr{8}{3}\pi \ , & \eta_4 = \fr{8}{3}\pi \ .
\end{array}
\eea

\section*{Appendix B: Quantization rules and the dibaryon isospin}

\renewcommand{\theequation}{B.\arabic{equation}}
\setcounter{equation}{0}

In case the background field is restricted to $SU(2)$ one finds that
in the bound state approach (as in the case of the normal $SU(2)$
Skyrme model) there is a quantization ambiguity on whether a baryon
number B configuration has to be quantized as a fermion or as a boson.
Furthermore one has no information on which $SU(3)$ multiplet a
quantized bound state system with $J$, $I$ and $S$ should belong. In
order to avoid these ambiguities in the bound state approach the
authors of Ref.\cite{SW91} suggested to introduce a third light
flavour - degenerated with the $u$ and $d$ flavours - called ``funny
strange" flavour and to embed the bound state ansatz (4) into
$SU(4)^{(f)}$ (where the label $(f)$ distinguishes this group from a
physical $SU(4)$ flavour group) as follows:
\be
U^{(f)} =
\left(\begin{array}{cc}\sqrt{U_\pi^{(f)}}  &0\\\ 0&1\end{array}\right)
 \ U_K^{(f)} \
\left(\begin{array}{cc}\sqrt{U_\pi^{(f)}}  &0\\\ 0&1\end{array}\right) \ .
\label{uf}
\ee
Now $U_\pi^{(f)}$ belongs to $SU(3)^{(f)}$,
\be
U_\pi^{(f)}  =
\left(\begin{array}{cc}u_\pi &0\\\ 0&1\end{array}\right) \ ,
\nonumber
\ee
and $U_K^{(f)}$ has in terms of $K^{(f)}$ the same form as $U_K$ in
terms of $K$, where
\be
K^{(f)} =
\left(\begin{array}{c}K \\ 0 \end{array}\right)
=
 \left(\begin{array}{c}K^+ \\ K^0 \\ 0\end{array}\right)
\nonumber.
\ee

After collective quantization of the $SU(3)^{(f)}$ flavour degrees of
freedom (u, d and f) the Wess-Zumino term \footnote{The derivation of
this result is in complete analogy to the one for the quantization of
an $SU(3)$ skyrmion. In that case, however, the Wess--Zumino term leads
to the constraint $Y_R = N_c \ B / 3$
\cite{BLRS85}.}  leads to a constraint on the ``funny" right-hypercharge
$Y_R^{(f)}$ \cite{SW91},
\be
Y_R^{(f)} = { N_c \ B + S \over 3} \ ,
\label{yrf}
\ee
where $N_c$ is the number of colours, $B$ the baryon charge of the
configuration and $S$ the physical ( {\it not} ``funny"~) strangeness.
The derivation of this relation assumes only that the background field
$u_\pi$ belongs to $SU(2)$ and that $K$ has the form (6). The result
(\ref{yrf}) is as well applicable for the $SU(3)^{(f)}$--extended
axial rotor system with $n \geq 2$ where $n=B$. For $B=2$ and $N_c =
3$ we have therefore the constraint
\be
Y_R^{(f)} = 2 + {S \over 3} \ .
\label{athree}
\ee
Of all possible $SU(3)^{(f)}$ multiplets with $Y_R^{(f)} $ given by
Eq.(\ref{athree}) only the ``minimal ones" (which are naturally the
energetically lowest ones) will interest us. ``Minimal" means that
they are composed in a minimal way out of $p$ $SU(3)^{(f)}$ triplet
and $q$ antitriplet representations. Therefore they have
\be
Y_R^{(f)} = {p + 2 q \over 3}   \ \ \ \ \ \  \mbox{and}  \ \ \ \ \ \  I_R^{(f)}
   = {p  \over 2} \ ,
\label{athreeb}
\ee
where $I_R^{(f)}$ is the right isospin.

The multiplets are then described by the $SU(3)^{(f)}$ rotation matrices
\be
D^{p,q}_{Y^{(f)} I^{(f)} I_3^{(f)} , Y_R^{(f)} I_R^{(f)} I_{R,z}^{(f)}}
\left( \gamma_1, \gamma_2 , ...,\gamma_8\right) \ ,
\label{afour}
\ee
where $\gamma_1, \gamma_2 ,\gamma_3$ are of course the physical Euler
angles for isospin rotations, $Y^{(f)}$ is the funny hypercharge of a
member of the multiplet, $I^{(f)}$ its isospin, $I_3^{(f)}$ the
projection of the isospin in the lab. frame and $I_{R,z}^{(f)}$ the
projection of the right isospin in the body-fixed frame.

Since the ``funny strangeness" was introduced only to serve as an
auxiliary quantity, the physical states should belong to those $SU(2)$
submultiplets of a given $SU(3)^{(f)}$ multiplet which do not have any
``funny strange" component. In practice this means that we only work
with the upper row of a given minimal $SU(3)^{(f)}$ multiplet. Thus we
have $Y=Y^{(f)}=Y_R^{(f)}$ and $I=I^{(f)}=I_R^{(f)}=p/2$ where $Y$ and
$I$ are the hypercharge and the isospin of the physical states. In
this case, the $SU(3)^{(f)}$ rotation matrix (\ref{afour}) simplifies
to
\be
D^{p,q}_{Y_R I I_3 , Y_R I I_3^{bf}}
\left( \gamma_1, \gamma_2 , ...,\gamma_8\right) \ ,
\label{afive}
\ee
which is equivalent, as the isospin content is concerned, to the usual
$SU(2)$ isospin rotation matrix
\be
D^{I}_{I_3 , I_3^{bf}} \ ,
\left( \gamma_1, \gamma_2 , \gamma_3\right)
\label{asix}
\ee
where $I_3$ is the isospin projection in the lab. frame and $I_3^{bf}$
the one in the body-fixed frame.

     From Eqs.(\ref{athreeb},\ref{athree}) we can now construct the
allowed funny $SU(3)^{(f)}$ multiplets for a bound state configuration
with $B=2$ and given $S$. For $S=0$ we find $(p,q)$ = (0,3), (2,2),
(4,1) and (6,0) which correspond to the usual 6
quark flavour multiplets: $\bar{10}$, $27$, $35$ and $28$. For $S=-1$,
we find $(p,q) = (1,2), \ (3,1) \ \mbox{and} \ (5,0)$ which correspond
to the $SU(3)^{(f)}$ multiplets: $\bar{15}$, $24$ and $21$
respectively. Note that these numbers are the same as for the 5 quark
$SU(3)$ multiplets which follow after removal of one quark from the
6 quark system.

     In Table \ref{rotorqn} all minimal ``funny" multiplets $(p,q)$
for the $B=2$ bound state configurations with strangeness $S$ between
0 and $-6$ are listed. From that table one can deduce that the states
characterized by the rotation matrix of the background field
(\ref{afive}) correspond in the quark model to those minimal quark
flavour configurations which involve only the $u$ and $d$ quarks of a
given $B=2$ state. E.g. for $S=-6$ the funny multiplet is a singlet
signalling that there is no $u$, $d$ content. The role of the strange
quarks (6 in the $S=-6$ system) is taken over by the kaons coupled to
the $n=2$ soliton background. Note that any $B=2$ state with $S \leq -
7$ or $S \geq 1$ corresponds to a non-minimal $SU(3)^{(f)}$ multiplet
for the rotor.

\pagebreak
\section*{Table captions}

\vspace{1.cm}

\indent {\bf Table 1:} The allowed quantum numbers in the $B=2$ bound
statesystem. In the rows the following quantities are listed: (1) the
strangeness $S$,
(2) the allowed values of isospin $I$,
(3) the magnitude of the isospin projection in
the body fixed frame, $|I_3^{bf}| \leq I$, (4) the magnitude of the
projection of the kaon grand-spin on the body fixed frame, $| T_3|
\leq |S/2|$, (5) the magnitude of the projection of the total angular
momentum on the body fixed frame, $|J_3^{bf}|= 2|I_3^{\bf}+T_3 |$ (see
Eq.(\ref{constraint})) with $|J_3^{bf}| \leq 2$, the parity $P$ as given by
Eq.(\ref{parity}), and the number of states with the same $J_3^{bf}$
quantum number, (6) the total angular momentum $J\leq 2$ and the
corresponding $SU(3)$ representations.

\vspace{1.cm}
\noindent {\bf Table 2:}
Numerical values of the parameters appearing in the dibaryon mass formulae
Eqs.(\ref{massformula}-\ref{mrotst}).
SET~A corresponds to the massless pion case while
SET~B corresponds to massive pions.

\vspace{1.cm}

\noindent {\bf Table 3:}
Rotational contributions to the dibaryon masses. Listed are those
corresponding to the allowed states with $M_{rot} \le 250 \ MeV$.
SET~A and B are as in Table 2. $NN$, $N\Lambda$ and $\Lambda \Lambda$
stand for the sum of the rotational contributions to the corresponding
particles and serve as rotational threshold in each particular channel.

\vspace{1.cm}

\noindent {\bf Table 4:}
Contributions (in $MeV$) to the mass of the H--particle given
relative to twice the corresponding contribution to the
$\Lambda$ mass.

\vspace{1.cm}

\noindent {\bf Table 5:}
Allowed ``funny" multiplets $(p,q)$ for the $B=2$ bound state
configurations.

\pagebreak

\centerline{{\Large \bf Table 1}}
\vspace{1.cm}

\begin{table}[h]
\label{rotorstates}
\centering
\begin{tabular}{|c|c|ll|l|l|}
\hline
  $S$  & $I$ & $|I_3^{bf}|$ & $|T_3|$ &
   $(|J_3^{bf}|)^{P}_{\rm deg}$ & $J^P$($SU(3)$ repr.) $\leq 2$  \\
\hline
  0 & 0 & 0       & 0 & $0^+_1$        & $1^+(\bar{10})$      \\
    & 1 & 0,1     & 0 & $0^+_1; 2^-_1$ & $0^+(27); 2^\pm(27)$ \\
    & 2 & 0,1,2   & 0 & $0^+_1; 2^-_1$ & $1^+(35); 2^-(35)$   \\
    & 3 & 0,1,2,3 & 0 & $0^+_1; 2^-_1$ & $0^+(28); 2^\pm(28)$\\[2.mm]
 -1 & \half     & \half                        & \half & $0^+_2; 2^-_1$ &
          $1^+(\bar{10}), 0^+(27); 2^\pm(27)$ \\[1.mm]
    & \threehalf& \half, \threehalf            & \half & $0^+_2; 2^-_2$ &
          $0^+(27), 1^+(35); 2^\pm(27), 2^-(35)$ \\[1.mm]
    & \fivehalf & \half, \threehalf, \fivehalf & \half & $0^+_2; 2^-_2$ &
          $0^+(28), 1^+(35); 2^\pm(28), 2^-(35)$ \\[2.mm]
 -2 & 0 & 0     & 0,1 & $0^+_1; 2^-_1$ &
          $0^+(27); 2^\pm(27)$ \\
    & 1 & 0,1   & 0,1 & $0^+_3; 2^-_2$ &
          $0^+(27), 1^+(35,\bar{10}); 2^\pm(27), 2^-(35)$ \\
    & 2 & 0,1,2 & 0,1 & $0^+_3; 2^-_3$ &
          $0^+(28,27), 1^+(35); 2^\pm(28,27), 2^-(35)$ \\[2.mm]
 -3 & \half     & \half             & \half, \threehalf & $0^+_2; 2^-_2$ &
          $0^+(27), 1^+(35); 2^\pm(27), 2^-(35)$ \\[1.mm]
    & \threehalf& \half, \threehalf & \half, \threehalf & $0^+_4; 2^-_3$ &
          $0^+(28,27), 1^+(35,\bar{10});
          2^\pm(28,27), 2^-(35)$ \\[2.mm]
 -4 & 0 & 0   & 0,1,2 & $0^+_1; 2^-_1$ &
          $1^+(35); 2^-(35)$ \\
    & 1 & 0,1 & 0,1,2 & $0^+_3; 2^-_3$ &
          $0^+(28,27), 1^+(35); 2^\pm(28,27), 2^-(35)$ \\[2.mm]
 -5 & \half & \half & \half, \threehalf, \fivehalf & $0^+_2; 2^-_2$ &
          $0^+(28), 1^+(35); 2^\pm(28), 2^-(35)$ \\[2.mm]
 -6 & 0 & 0 & 0,1,2,3 & $0^+_1; 2^-_1$ &
          $0^+(28); 2^\pm(28)$ \\[2.mm]
\hline
\end{tabular}
\end{table}

\vspace{1.5cm}
\begin{center}
{\Large \bf Table 2}

\vspace{1.cm}
\begin{tabular}{|c|c|c|c|c|c|c|c|}
\hline
 &$M_{sol}$ & $\caa$ & $\cab$ & $\cac$ & $\veps$ & $c_1$ & $c_2$ \\ \hline
 & $MeV$     &  $fm$    &  $fm$ &  $fm$     &  $MeV$   &       &      \\ \hline
SET A & 1675 &  2.62    &  1.75 &  1.17     &   238    & .623  & .436 \\ \hline
SET B & 1675 &  3.22    &  2.11 &  1.42     &  226     & .554  & .334 \\ \hline
\end{tabular}
\end{center}

\pagebreak

\begin{center}
{\Large \bf Table 3}

\vspace{1.cm}
\begin{tabular}{|c|c|c|c|c|}\hline
    &      & \multicolumn{2}{|c|}{ This model } & Ref.\cite{KM88} \\
\cline{3-4}
    & (I,J$^\pi$) & SET A & SET B &     \\
\hline
     & $0,1^+$ &  75 & 61  & 76 \\
 S=0 & $1,0^+$ & 113 & 93  & 120 \\
     & NN      & 147 & 147 & 147 \\
     & $1,2^-$ & 216 & 177 & 212 \\ \hline
         & $1/2,0^+$ & 61  &  46 & 76  \\
         & $1/2,1^+$ & 87  &  75 & 83  \\
 $S=-1$  & N$\Lambda$& 92  &  85 & 90  \\
         & $3/2,0^+$ & 157 & 138 & 160 \\
         & $1/2,2^-$ & 165 & 129 & 169 \\ \hline
         & $0,0^+$ &  21  &  11  & 39   \\
  &$\Lambda\Lambda$&  37  &  22  & 33   \\
         & $1,0^+$ &  79  &  66  & 88   \\
 $S=-2$  & $1,1^+$ & 107  &  98  & 115  \\
         & $0,2^-$ & 119  &  88  & 130  \\
         & $1,2^-$ & 187  & 152  & 182  \\
         & $0,2^+$ & 247  & 195  & higher \\ \hline
\end{tabular}
\end{center}

\vspace{3.cm}

\begin{center}
{\Large \bf Table 4}

\vspace{1.cm}
\begin{tabular}{|c|r|r|}\hline
                      & SET A & SET B \\
\hline
$\Delta M_{sol}$          &  -52  & -54  \\
$2 \Delta \veps$          &   32 &   32  \\
$\Delta M_{rot}$          &  -15  & -12  \\ \hline
$M(H) - 2 M(\Lambda)$     &  -34 & -34  \\ \hline
\end{tabular}
\end{center}

\pagebreak

\vspace*{3.cm}

\centerline{{\Large \bf Table 5}}
\vspace{1.cm}

\begin{table}[h]
\label{rotorqn}
\centering
\begin{tabular}{|c|c|cc|c|c|}
\hline
  $S$  & $Y_R^{(f)}$ & p & q & $SU(3)^{(f)}$ repres.& $I=I_R$ \\
\hline
    0       &   2  &   0      &  3  & $\bar{10}$  & 0 \\
             &       &   2      &  2  &       27  & 1      \\
             &       &   4      &  1  &       35  & 2      \\
             &       &   6      &  0  &       28  & 3      \\[2.mm]
    -1     & 5/3&   1      &  2  & $\bar{15}$     & 1/2 \\
             &       &   3      &  1  &       24  & 3/2 \\
             &       &   5      &  0  &       21  & 5/2 \\[2.mm]
    -2     & 4/3&   0      &  2  & $\bar{6}$      &  0 \\
             &       &   2      &  1  &       15  &  1   \\
             &       &   4      &  0  &       15  &  2  \\[2.mm]
    -3     & 1    &   1      &  1  &         8    & 1/2    \\
             &       &   3      &  0  &       10  & 3/2      \\[2.mm]
    -4     & 2/3&   0      &  1  &  $\bar{3}$     & 0 \\
             &       &   2      &  0  &        6  & 1 \\[2.mm]
    -5     & 1/3&   1      &  0  &        3       & 1/2 \\[2.mm]
    -6     &    0 &   0      &  0  &        1     & 0   \\
 \hline
\end{tabular}
\end{table}

\end{document}